\documentclass[onecolumn,amsmath,amssymb,superscriptaddress,nofootinbib,11pt,a4paper]{revtex4}

\pdfoutput=1
\usepackage[T1]{fontenc}
\usepackage{CJK}
\usepackage{xcolor}
\usepackage{amsfonts}
\usepackage{epstopdf}
\usepackage{wrapfig} 
\usepackage{subcaption}
\usepackage{graphicx}  
\usepackage{dcolumn}   
\usepackage{bm}
\usepackage{float}
\usepackage[T1]{fontenc}
\usepackage{CJK}
\usepackage{xcolor}
\usepackage{amsfonts}
\usepackage{epstopdf}
\usepackage{wrapfig} 
\usepackage{subcaption} 
\usepackage{graphicx}  
\usepackage{dcolumn}   
\usepackage{bm}
\usepackage{float}
\usepackage[utf8]{inputenc}
\usepackage[T1]{fontenc}

\begin{document}

\title{Energy Extraction From the Kerr-Bertotti-Robinson Black Hole via Magnetic Reconnection under Circular Plasma and Plunging Plasma}

\author{Xiao-Xiong Zeng}
\email{xxzengphysics@163.com}
\affiliation{College of Physics and Electronic Engineering, Chongqing Normal University, \\Chongqing 401331, China}

\author{Ke Wang}
\email{kkwwang2025@163.com}
\affiliation{School of Material Science and Engineering, Chongqing Jiaotong University, \\Chongqing 400074, China}

\begin{abstract}
{ Recently, a class of exact solutions describing rotating black holes immersed in a uniform magnetic field has been proposed, and  various properties of such black holes remain unclear. This paper aims to explore how to extract energy based on the magnetic reconnection mechanism  in both the circular orbit region and the plunging region. After introducing the properties of this spacetime, we analyze physical quantities such as the size of the ergoregion, the event horizon, and the boundaries of the ergosphere. We then analyze the magnetic reconnection process within circular orbits. We  plot energy extraction parameter diagrams, and analyze the power and efficiency of energy extraction. The results show  that the magnetic field impedes energy extraction. Comparing with Kerr black holes and Kerr-Melvin black holes, we find that this black hole's energy extraction capability is stronger than that of the Kerr-Melvin black hole but weaker than that of the Kerr black hole. Furthermore, we study magnetic reconnection and energy extraction in the plunging region of this black hole. The results indicate that the magnetic field still impedes energy extraction. The black hole's energy extraction capability in this region also remains weaker than that of a Kerr black hole but stronger than that of a Kerr-Melvin black hole. Comparing the plunging region with the circular orbit region, we find that the energy extraction power in the plunging region is consistently stronger than in the circular orbit region.}
\end{abstract}

\maketitle
 \newpage
\section{Introdution}
Black holes are one of the most mysterious celestial objects in the universe. Since the detection of gravitational waves by Laser Interferometer Gravitational-wave Observatory (LIGO) \cite{41} and the imaging of the black hole shadow by the Event Horizon Telescope (EHT) \cite{42}, physicists have gained a deeper understanding of black holes. Previous research has demonstrated that matter surrounding a black hole significantly influences the trajectories of particles and the appearance of its shadow image. This is particularly evident in the EHT's published analyses of polarized imaging around the supermassive black hole M87* \cite{43} and the polarized imaging of the Galactic center black hole Sagittarius A* \cite{44}, which confirmed the presence of magnetic fields around these black holes. Therefore, studying the properties of magnetized black holes is  important both theoretically and astronomically. Currently, numerous studies have explored the physical properties of black holes immersed in magnetic fields. Most of these build upon the Schwarzschild-Melvin black hole solution, which has subsequently been generalized to include rotation, electric charge, and other parameters \cite{14,15,16,17,18,19,20,21,22,23,7,8,9,11,Barrientos:2024pkt}. However, these spacetimes have some drawbacks. For instance, the magnetic field decreases far away from the black hole, the geodesics cannot escape to infinity \cite{1}, and the algebraic type of the spacetime is $I$ \cite{23}. Recently, a new class of exact black hole solutions immersed in a magnetic field has been discovered \cite{1}. This black hole is referred to as the Kerr-Bertotti-Robinson black hole. Crucially, test particles can, in principle, escape to spatial infinity within this spacetime, and its algebraic type is $D$ \cite{1}. Type $D$ spacetimes often possess additional symmetries (Killing vector), which typically allow for the separation of variables in null geodesic equations. This significantly simplifies both the precise analytical and numerical studies of particle orbits. In contrast, type I spacetimes generally lack such symmetries, resulting in the inability to separate variables in the geodesic equations. Many properties of this new black hole remain unclear. This paper aims to explore how energy can be extracted from this black hole via magnetic reconnection in  the circular orbit region and  plunging region.

Extracting energy from black holes has long been a topic of great interest to physicists. Penrose discovered in 1969 that energy could be extracted from rotating black holes, a process now known as the Penrose process \cite{5}. A particle of energy $E$ falls freely from infinity into a black hole's ergosphere and splits. From an observer at infinity's viewpoint, a negative-energy fragment enters the event horizon, allowing a positive-energy fragment to escape along a geodesic. This constitutes energy extraction. Later, Wald pointed out \cite{24} that realizing the Penrose process in practice is extremely difficult. Building on Penrose's pioneering work, many physicists have discovered other mechanisms for energy extraction, such as superradiant scattering \cite{25}, the Blandford-Znajek process \cite{26}, among others. Comisso and Asenjo also discovered a highly efficient mechanism for energy extraction by employing the method of magnetic reconnection \cite{3} (for related earlier work, see \cite{45,46}). Furthermore, in some cases, its power exceeds that of the Blandford-Znajek process. Consequently, this work has been rapidly extended to other rotating black holes \cite{2,6,27,28,29,30,31,32,33,34,35,36,37,Fan:2024rsa,Fan:2024fcy}. Currently, research on the magnetic reconnection mechanism primarily focuses on the circular orbit region. Recently, reference \cite{13} proposed studying magnetic reconnection and energy extraction in the plunging region. Specifically, plasma initially performs circular motion outside the innermost stable circular orbit (ISCO). However, since circular orbits within the ISCO are unstable, the plasma plunges inward starting from the ISCO. References \cite{13,38,39,40} have already investigated magnetic reconnection and energy extraction in the plunging region. The results demonstrate that energy extraction from the black hole is indeed possible in the plunging region. More importantly, for moderate to low black hole spin values $a$ (where a significant plunging region exists inside the ISCO), both the power and efficiency of energy extraction in this plunging region are typically higher than those achievable from circular orbits in the equatorial plane.

Indeed, energy extraction from black holes immersed in a magnetic field has already been studied in reference \cite{6}. However, the spacetime used there was the Kerr-Melvin black hole, whose algebraic type is $I$. In contrast, this paper studies a new exact solution with algebraic type $D$. Furthermore, the magnetic reconnection investigated in \cite{6} was confined to the circular orbit region and did not consider the plunging region. Additionally, they analyzed only a moderate value of $B$ \cite{6}, without examining the energy extraction scenario near an extremal black hole. This paper will investigate magnetic reconnection and the resulting power and efficiency of energy extraction for the Kerr-Bertotti-Robinson black hole, in both the circular orbit region and the plunging region. We will also compare our findings with those for the Kerr-Melvin black hole. In the circular orbit region, the Kerr-Bertotti-Robinson black hole exhibits higher energy extraction power than the Kerr-Melvin black hole but lower power than the Kerr black hole, with energy extraction being more readily facilitated by either a moderate $B$ value or scenarios approaching an extremal black hole. In the plunging region, its energy extraction capability remains stronger than Kerr-Melvin's but weaker than Kerr's, with the magnetic field hampering extraction in both regions.

The remainder of this paper is organized as follows. In Section 2, we introduce the Kerr-Bertotti-Robinson spacetime and its key features. Section 3.1 presents the magnetic reconnection process within the circular orbit region. Section 3.2 analyzes the power and efficiency of energy extraction in the circular orbit region. Section 4.1 contributes to  the magnetic reconnection process in the plunging region. Section 4.2 conducts the efficiency and power analysis for the plunging region. Finally, we present our conclusions in Section 5.

\section{Introduction to Kerr-Bertotti-Robinson Spacetime}
In Boyer-Lindquist (BL) coordinates  with geometric units ($c = G = 1$), the metric expression for the Kerr-Bertotti-Robinson spacetime is given by \cite{1,52,53}
\begin{align}
\mathrm{d}s^{2}=\frac{1}{\Omega^{2}}\Big{[}-\frac{Q}{\rho^{2}} \big{(}\mathrm{d}t-a\sin^{2}\theta\,\mathrm{d}\varphi\big{)}^{2}+\frac{\rho^{2 }}{Q}\,\mathrm{d}r^{2}+\frac{\rho^{2}}{P}\,\mathrm{d}\theta^{2} +\frac{P}{\rho^{2}}\sin^{2}\theta\,\big{(}a\,\mathrm{d}t-(r^{2}+a^ {2})\,\mathrm{d}\varphi\big{)}^{2}\Big{]},\label{1}    
\end{align}
where
\begin{align}
\rho^{2}=r^{2}+a^{2}\cos^{2}\theta,P=1+B^{2}\Big{(}m^{2}\,\frac{I_{2}}{I_{1}^{2}}-a^{2}\Big{)}\cos^{2 }\theta,Q=\,\big{(}1+B^{2}r^{2}\big{)}\,\Delta,
\end{align}
\begin{align}
\Omega^{2}=\,\big{(}1+B^{2}r^{2}\big{)}-B^{2}\Delta\cos^{2}\theta,
\Delta=\Big{(}1-B^{2}m^{2}\,\frac{I_{2}}{I_{1}^{2}}\Big{)}r^{2}-2m \,\frac{I_{2}}{I_{1}}\,r+a^{2},
\end{align}
\begin{align}
I_{1}=1-\tfrac{1}{2}B^{2}a^{2}, I_{2}=1-B^{2}a^{2}.   
\end{align}
The corresponding metric components can be expressed as
\begin{align}
g_{tt} = \frac{1}{\Omega^2 \rho^2} ( -Q + P a^2 \sin^2 \theta ),g_{t\varphi} = g_{\varphi t} = \frac{1}{\Omega^2 \rho^2} a \sin^2 \theta ( Q - P (r^2 + a^2) ),
\end{align}
\begin{align}
g_{rr} = \frac{1}{\Omega^2} \frac{\rho^2}{Q}, g_{\theta\theta} = \frac{1}{\Omega^2} \frac{\rho^2}{P}, g_{\varphi\varphi} = \frac{1}{\Omega^2 \rho^2} \sin^2 \theta \left( -Q a^2 \sin^2 \theta + P (r^2 + a^2)^2 \right). 
\end{align}
The inverse of the metric components is
\begin{align}
g^{tt} &= -\dfrac{\Omega^{2}}{P Q \rho^{2}} \left[ P (r^{2} + a^{2})^{2} - Q a^{2} \sin^{2} \theta \right],
g^{t\varphi} = g^{\varphi t} = \dfrac{\Omega^{2} a \left( Q - P (r^{2} + a^{2}) \right)}{P Q \rho^{2}} , 
\end{align}
\begin{align}
g^{\varphi\varphi}= \dfrac{\Omega^{2} \left( Q - P a^{2} \sin^{2} \theta \right)}{P Q \rho^{2} \sin^{2} \theta},
g^{rr} = \dfrac{\Omega^{2} Q}{\rho^{2}},g^{\theta\theta}= \dfrac{\Omega^{2} P}{\rho^{2}}.     
\end{align}
Here, $m$ is the mass, $a$ is the black hole spin, and $B$ is the magnetic field strength. When $B = 0$, the metric reduces to the Kerr metric, when $m = 0$, it reduces to the Bertotti-Robinson metric, and when $a = 0$, it reduces to the Schwarzschild-Bertotti-Robinson metric.

\begin{figure}[!h]
\centering
\includegraphics[width=0.9\linewidth]{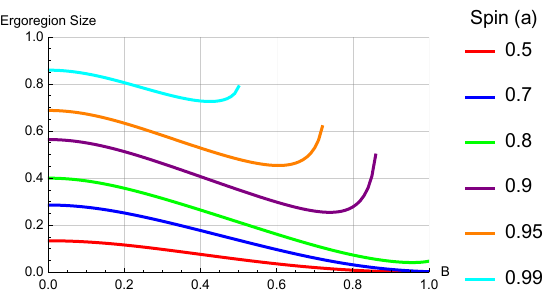}
\caption{ Variation of the size of the ergoregion versus  $B$ for different values of $a$.}
\label{fig:1}
\end{figure}

We follow reference \cite{3}, where a simplifying assumption is made that the magnetic reconnection process is modeled in the current sheet located on the equatorial plane. Therefore, we restrict the particle motion to the equatorial plane. While this approach may limit the generality of our conclusions, reference \cite{54} shows that the geodesic equations of this spacetime cannot be solved via variable separation for massive particles. This significantly increases the complexity of analyzing magnetic reconnection under general conditions. So, we consider the equations of motion for particles confined to the equatorial plane ($\theta = \pi/2$). The normalization condition for the geodesic equations is given by
\begin{align}
g_{\mu\nu} \frac{dx^\mu}{d\tau} \frac{dx^\nu}{d\tau} = -\varepsilon,\label{9}    
\end{align}
for photons, $\varepsilon = 0$, and for massive particles, $\varepsilon = 1$, where $\tau$ is the affine parameter. This metric possesses two conserved quantities, the energy $E$ and the angular momentum $L$, which can be expressed as
\begin{align}
p_t = g_{tt} \frac{dt}{d\tau} + g_{t\varphi} \frac{d\varphi}{d\tau} = -E, \quad p_\varphi = g_{\varphi t} \frac{dt}{d\tau} + g_{\varphi\varphi} \frac{d\varphi}{d\tau} = L, \label{10}
\end{align}
where
\begin{align}
\frac{dx^\mu}{d\tau} = g^{\mu\nu} p_\nu.    
\end{align}
For $\mu = t$
\begin{align}
\frac{dt}{d\tau} = g^{tt} p_t + g^{t\varphi} p_\varphi = g^{tt} (-E) + g^{t\varphi} L.\label{12}     
\end{align}
For $\mu = \varphi$
\begin{align}
\frac{d\varphi}{d\tau} = g^{\varphi t} p_t + g^{\varphi \varphi} p_\varphi = g^{\varphi t} (-E) + g^{\varphi \varphi} L. \label{13}    
\end{align}
In the equatorial plane ($\theta = \pi/2$), we have
\begin{align}
d\theta/d\tau = 0.\label{14} 
\end{align}
Substituting \eqref{12}, \eqref{13}, and \eqref{14} into \eqref{9}, and utilizing the conserved quantities \eqref{10} and the corresponding metric components, yields the radial equation
\begin{align}
\left( \frac{dr}{d\tau} \right)^2 = \frac{ \Omega^4 }{ r^4 } \left( \left[E (r^2 + a^2) - a L\right]^2 - \Omega^2 (a E - L)^2 \Delta - \varepsilon \Delta r^{2} \right).\label{15}
\end{align}
Equations \eqref{12}, \eqref{13}, \eqref{14}, and \eqref{15} constitute the geodesic equations for particles in the equatorial plane. According to \eqref{1}, the horizon is located at $Q=0$. This equation has at most two real roots; the larger root corresponds to the event horizon, while the smaller root corresponds to the inner horizon. The event horizon can be expressed as
\begin{align}
r_{+}=\frac{m\,I_{2}+\sqrt{m^{2}I_{2}-a^{2}I_{1}^{2}}}{I_{1}^{2}-B^{2}m^{2}I_{2}}\,I_{1}\,.  
\end{align}
In this paper, we consider unit mass, setting $m=1$.  The boundary of ergosphere is located at $g_{tt}=0$. We plot the size of the ergoregion\footnote{It is defined as the difference between the value at the boundary of ergosphere  and the value at the event horizon. } as a function of $B$ for different values of $a$, as shown in Figure \ref{fig:1}. Curves for higher spins, such as $a=0.99$, terminate when they reach the maximum allowed spin at $B=0.5$. Figure \ref{fig:1} shows that the size of the ergoregion increases with increasing spin parameter $a$, which facilitates energy extraction. For smaller spins, such as $a=0.5$ and $a=0.7$, the ergoregion size decreases as $B$ increases, which hinders energy extraction. For larger spins, the ergoregion size first decreases and then increases with increasing $B$. The subsequent increase is due to effects near the extremal black hole, which, as will be seen later, also facilitates energy extraction.

\begin{figure}[!h]
\centering
\includegraphics[width=0.9\linewidth]{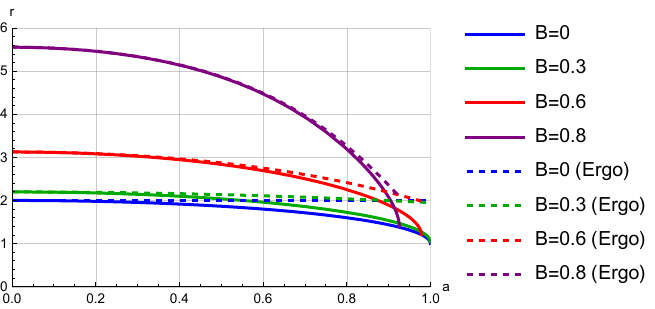}
\caption{ Positions of the event horizon and boundary of ergosphere versus  $a$ for different values of $B$.}
\label{fig:2}
\end{figure}

We also plot the event horizon and the boundary of ergosphere as functions of $a$ for different $B$ values, shown in Figure \ref{fig:2}. Solid curves represent the event horizon, while dashed curves represent the boundary of ergosphere.
From Fig. \ref{fig:2}, it can be seen that as $B$ increases, the positions of the event horizon and boundary of ergosphere shift outward. The boundary of ergosphere becomes more curved, and the event horizon and boundary of ergosphere move closer together, separating only slightly near the extremal black hole. This explains why the size of the ergoregion increases near the extremal black hole case in Fig. \ref{fig:1}.

Considering particles undergoing circular orbital motion in the equatorial plane, circular orbits for these particles can exist from infinity down to the photon sphere. The photon sphere radius satisfies \cite{Zeng1,Zeng2}
\begin{align}
R(r)=0, R^{\prime}(r)=0.    
\end{align}
According to \eqref{15}, $R(r)$ can be written as
\begin{align}
 R(r)= \left[E (r^2 + a^2) - a L\right]^2 - \Omega^2 (a E - L)^2 \Delta - \varepsilon \Delta r^{2}.    
\end{align}
For the ISCO, the conditions are
\begin{align}
R(r)=0, R^{\prime}(r)=0,R^{\prime\prime}(r)=0. \label{19}
\end{align}
For retrograde orbits, the photon sphere radius lies outside the ergosphere, making energy extraction impossible under circular orbits. Regarding the plunging region, although particles can enter the ergosphere from the ISCO, the energy density per enthalpy at infinity for both accelerated and decelerated plasma remains greater than zero, which still prevents energy extraction \cite{13}. Therefore, for energy extraction, only prograde orbits need to be considered. The Keplerian angular velocity for particles in prograde orbits is given by \cite{2}
\begin{align}
\Omega=\frac{d \varphi / d \tau}{d t / d \tau}= \frac{ - g_{t\varphi,r} + \sqrt{ \left( g_{t\varphi,r} \right)^2 - g_{tt,r}  g_{\varphi\varphi,r} } }{ g_{\varphi\varphi,r} }.\label{20}    
\end{align}
The expressions for energy $E$ and angular momentum $L$ are \cite{2} 
\begin{align}
E = - \frac{ g_{t t} + g_{t\varphi} \Omega }{ \sqrt{ -g_{t t} - 2g_{t\varphi} \Omega - g_{\varphi\varphi} \Omega^{2} } },L = \frac{ g_{t\varphi} + g_{\varphi\varphi} \Omega }{ \sqrt{ -g_{t t} - 2g_{t\varphi} \Omega - g_{\varphi\varphi} \Omega^{2} } }.\label{21}
\end{align}
For particles  satisfying the circular orbit conditions, equations \eqref{20} and \eqref{21} must yield real values. This requires that the expression under the square root in \eqref{20} and \eqref{21} be greater than zero.

\section{Energy Extraction from Kerr-Bertotti-Robinson Black Holes in Circular Orbital Regions}
\subsection{Magnetic Reconnection Process in Circular Orbits}
The magnetic reconnection process follows the Comisso-Asenjo framework \cite{3} for equatorial orbits. We first review the Comisso-Asenjo process.
Observations are conducted using the Zero Angular Momentum Observer (ZAMO) reference frame \cite{4}. In this frame, the spacetime interval is expressed as  
\begin{align}
ds^{2} = -d\hat{t}^{2} + \sum_{i=1}^{3} (d\hat{x}^{i})^{2} = \eta_{\alpha\beta}  d\hat{x}^{\alpha} d\hat{x}^{\beta}, 
\end{align}
with coordinate transformations defined by
\begin{align}
 d \hat{x^{i}}=\sqrt{g_{i i}} d x^{i}-\alpha \beta^{i} d t,d\hat{t}=\alpha d t,
\end{align}
where the lapse function $\alpha$, shift vector components $\beta$, and angular velocity $\omega^{\varphi}$ are given by
\begin{align}
\omega^{\varphi}=\frac{-g_{t\varphi }}{g_{\varphi \varphi}},\alpha=\sqrt{\left(-g_{t t}+\frac{g_{\varphi t}^{2}}{g_{\varphi \varphi}}\right)}, \beta^{\varphi}=\frac{\sqrt{g_{\varphi \varphi}} \omega^{\varphi}}{\alpha}, \beta^{r}=\beta^{\theta}=0. 
\end{align}
Within the ZAMO frame, the Keplerian orbital velocity is defined as
\begin{align}
\hat{v}_{K}=\frac{1}{\alpha}\left(\sqrt{g_{\varphi \varphi}} \Omega-\alpha \beta^{\varphi}\right). 
\end{align}
Our analysis employs the single-fluid plasma description, characterized by the energy-momentum tensor
\begin{align}
T^{\mu\nu}=p g^{\mu \nu}+w u^{\mu} u^{\nu}+F^{\mu}{}_{\sigma} F^{\nu \sigma}-(1 / 4) g^{\mu \nu} F^{\alpha \beta} F_{\alpha \beta}, 
\end{align}
where $w$ denotes enthalpy density, $p$ is pressure, $u^{\mu}$ represents the four-velocity, and $F_{\alpha\beta}$ is the electromagnetic field tensor. The magnetic reconnection is assumed to be highly efficient 
\footnote{In this case, magnetic energy undergoes near-total conversion to kinetic energy.}, in which contributions from the electromagnetic field tensor become negligible. Leveraging the plasma's adiabatic and incompressible properties, the energy density per enthalpy at infinity for the accelerated and decelerated plasma streams is approximated by \cite{3}
\begin{align}
e_{\pm}^{\infty}=\alpha \hat{\gamma}_{K}\left[\left(1+\beta^{\varphi} \hat{v}_{K}\right)\sqrt{1+\sigma} \pm \cos \xi\left(\beta^{\varphi}+\hat{v}_{K}\right) \sqrt{\sigma} - \frac{1}{4} \frac{\sqrt{1+\sigma} \mp \cos{\xi} \hat{v}_{K} \sqrt{\sigma}}{\hat{\gamma}_{K}^{2}\left(1+\sigma-\cos ^{2}{\xi} \hat{v}_{K}^{2} \sigma\right)}\right]. \label{27}
\end{align}
Here, $\xi$ corresponds to the fluid's azimuthal angle in the local rest frame, while $\sigma$ signifies the plasma magnetization parameter, which can be expressed as 
\begin{align}
\sigma=B^2/w,\label{28}
\end{align}
$\hat{\gamma}_{K}$ is the Lorentz factor of $\hat{v}_{K}$ , which can be expressed as 
\begin{align}
\hat{\gamma}_{K} = \frac{1}{\sqrt{1 - \hat{v}_{K}^{2}}}.
\end{align}
For a relativistically hot plasma characterized by the equation of state $w = 4p$, energy extraction, which is analogous to the Penrose particle-splitting mechanism \cite{5}, requires the simultaneous satisfaction of two criteria
\begin{align}
e_{-}^{\infty}<0, \Delta e_{+}^{\infty}=e_{+}^{\infty}-\left[1-\frac{p\Gamma}{w(\Gamma-1)}\right]=e_{+}^{\infty}>0, 
\end{align}
here, $\Gamma$ denotes the polytropic index, which is treated  as $\Gamma = 4/3$ here.

\begin{figure}[!h]
  \centering
  \begin{subfigure}{0.6\textwidth}
    \centering
    \includegraphics[width=\linewidth]{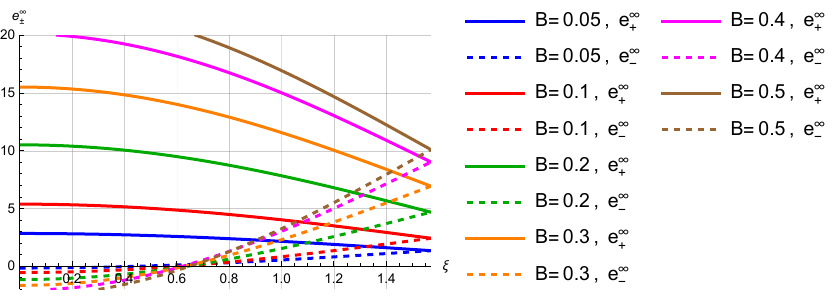}
    \caption{} 
    \label{fig:3a}
  \end{subfigure}
  \begin{subfigure}{0.37\textwidth}
    \centering
    \includegraphics[width=\linewidth]{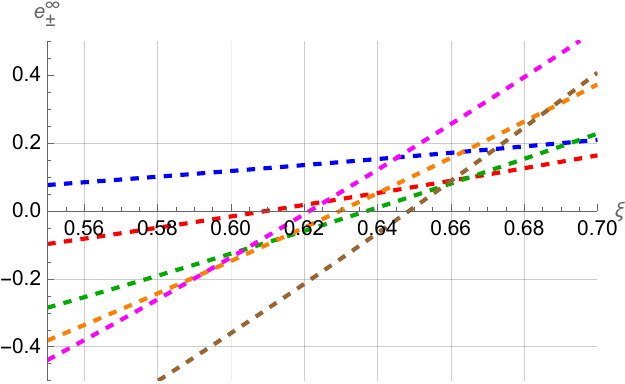} 
    \caption{} 
    \label{fig:3b}
  \end{subfigure}
  \caption{Variation of $e_{+}^{\infty}$ and $e_{-}^{\infty}$ with azimuthal angle $\xi$ for different values of $B$.}
  \label{fig:3}
\end{figure}

We plot the variation of $e_{+}^{\infty}$ and $e_{-}^{\infty}$ with azimuthal angle $\xi$ for different values of $B$ in Fig. \ref{fig:3}. Here $r$ denotes the dominant reconnection radial location. As explained in \cite{3}, the equatorial current sheet is unstable to the plasmoid instability \cite{55,56}, leading to fragmentation into multiple X-points. Among these, the dominant reconnection X-point is the one located at the intersection of the separatrices that encompass the global reconnection flow. This is the point that determines the reconnection dynamics and is the one referred to by the distance $r$. Without loss of generality, we set $a=0.99$, $w=0.001$, and $r=1.5$. Solid lines represent $e_{+}^{\infty}$, dashed lines represent $e_{-}^{\infty}$, with colors blue, red, green, orange, magenta, and brown corresponding to $B=0.05, 0.1, 0.2, 0.3, 0.4, 0.5$ respectively. The right panel shows a zoomed-in view of the left panel near $e_{-}^{\infty}=0$.
\begin{figure}[!h]
  \centering
  \begin{subfigure}{0.32\textwidth}
    \centering
    \includegraphics[width=\linewidth]{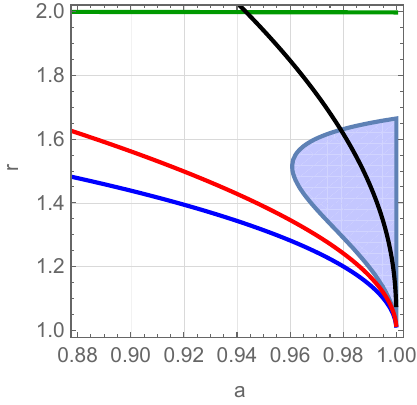}
    \caption{$B$=0.05} 
    \label{fig:4a}
  \end{subfigure}
  \begin{subfigure}{0.32\textwidth}
    \centering
    \includegraphics[width=\linewidth]{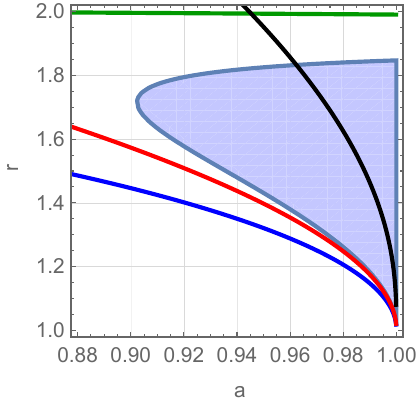} 
    \caption{$B$=0.1} 
    \label{fig:4b}
  \end{subfigure}
 \begin{subfigure}{0.33\textwidth}
    \centering
    \includegraphics[width=\linewidth]{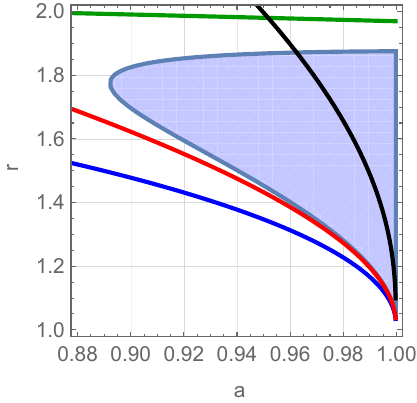} 
    \caption{$B$=0.2} 
    \label{fig:4c}
  \end{subfigure}
 \begin{subfigure}{0.32\textwidth}
    \centering
    \includegraphics[width=\linewidth]{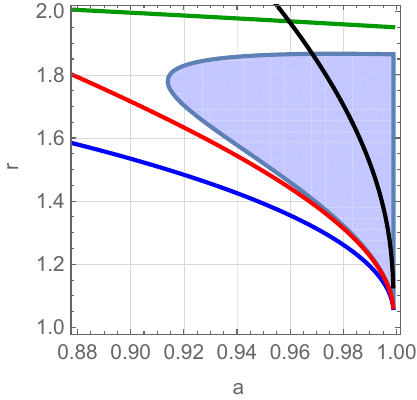} 
    \caption{$B$=0.3} 
    \label{fig:4d}
  \end{subfigure}
 \begin{subfigure}{0.32\textwidth}
    \centering
    \includegraphics[width=\linewidth]{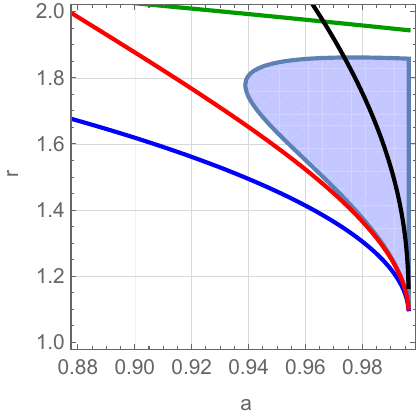} 
    \caption{$B$=0.4} 
    \label{fig:4e}
  \end{subfigure}
 \begin{subfigure}{0.32\textwidth}
    \centering
    \includegraphics[width=\linewidth]{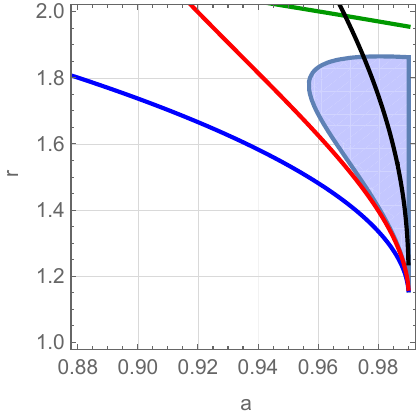} 
    \caption{$B$=0.5} 
    \label{fig:4f}
  \end{subfigure} 
  \begin{subfigure}{0.32\textwidth}
    \centering
    \includegraphics[width=\linewidth]{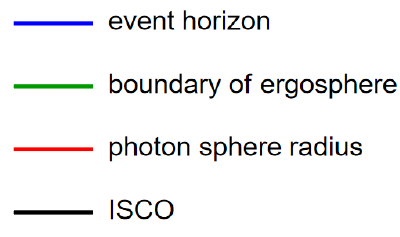} 
    \caption{Legend} 
    \label{fig:4f}
  \end{subfigure} 
  \caption{Energy extraction allowable regions for different values of $B$. }
  \label{fig:4}
\end{figure}
From (a)  in Fig. \ref{fig:3}, it can be observed that $e_{+}^{\infty}$ is always greater than zero. Therefore, for energy extraction to occur, the key condition is whether $e_{-}^{\infty}$ is less than zero. It is also evident that $e_{-}^{\infty}$ becomes negative only for smaller values of the azimuthal angle $\xi$, enabling energy extraction. This behavior aligns with the findings in Ref. \cite{3}. Additionally, at smaller $\xi$, as $B$ increases, $e_{-}^{\infty}$ decreases further. This effect arises from \eqref{28}, where  larger $B$ values correspond to higher $\sigma$. The critical value where $e_{-}^{\infty}$ equals zero is denoted as $\varpi$. As shown in (b) in Fig. \ref{fig:3}, $\varpi$ initially increases with $B$, then decreases, and finally increases again near the extremal black hole. This indicates that either a moderate $B$ value or proximity to an extremal black hole is more favorable for energy extraction.

Next, we plot the parameter space for energy extraction in the $r$-$a$ plane, specifically the region where $e_{-}^{\infty} < 0$, as shown in Fig. \ref{fig:4}. Here, blue, red, black, and green curves represent the event horizon, photon sphere radius, ISCO, and boundary of ergosphere, respectively. We set $w=0.001$ and $\xi=\pi/12$.
Fig. \ref{fig:4} demonstrates that as $B$ increases, the area of the allowed energy extraction  region first expands and then contracts. This again confirms that a moderate $B$ value is more conducive to energy extraction.
\subsection{Power and Efficiency of Energy Extraction in Circular Orbits}
\begin{figure}[!h]
  \centering
  \begin{subfigure}{0.49\textwidth}
    \centering
    \includegraphics[width=\linewidth]{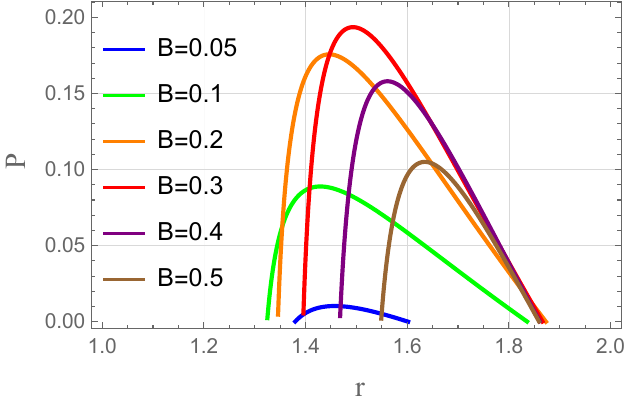}
    \caption{a=0.97} 
    \label{fig:5a}
  \end{subfigure}
  \begin{subfigure}{0.49\textwidth}
    \centering
    \includegraphics[width=\linewidth]{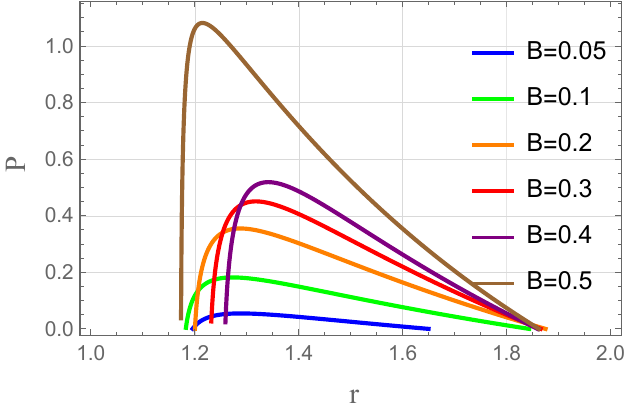} 
    \caption{a=0.99} 
    \label{fig:5b}
  \end{subfigure}
  \caption{Variation of energy extraction per unit enthalpy power versus  $r$ for different values of $B$.}
  \label{fig:5}
\end{figure}

Having confirmed the feasibility of energy extraction, we proceed to compare its power and efficiency. The per unit enthalpy power for energy extraction is given by \cite{3}
\begin{align}
P=-e^{\infty}_- A_{i n} U_{i n}.\label{31}
\end{align}
For collisionless conditions,  $U_{i n} \approx 0.1$ \cite{57}, for collisional conditions,  $U_{i n} \approx 0.01$ \cite{58}. This work considers the collisionless case. $A_{i n}$ is the cross-sectional area of the inflowing plasma can be expressed  as
\begin{align}
A_{i n} \sim\left(r_{E}^{2}-r_{p }^{2}\right), \label{32}
\end{align}
where $r_{E}$ is the boundary of ergosphere and $r_{p }$ is the photon sphere radius.
We plot the variation of energy extraction per unit enthalpy power with $r$ for different values of $B$ in Fig. \ref{fig:5}, setting $\xi=\pi/12$ and $w=0.001$. The left panel corresponds to $a=0.97$ while the right panel shows $a=0.99$. Blue, green, orange, red, purple, and brown curves represent $B=0.05, 0.1, 0.2, 0.3, 0.4, 0.5$ respectively.

From Fig. \ref{fig:5}, it can be observed that the power initially increases and then decreases with $r$. The left panel shows that away from extremal black holes, as $B$ increases, the power first rises and then falls, again indicating that a moderate $B$ value is optimal for energy extraction. The right panel reveals that near extremal black holes, the power increases monotonically with $B$, exceeding the values in the left panel. This enhancement stems from the higher spin parameter in the right panel, demonstrating that proximity to extremal black holes significantly facilitates energy extraction. Synthesizing Fig. \ref{fig:5}, we conclude that for non-extremal black holes, a moderate $B$ value is suitable for energy extraction, whereas near extremal black holes, the power increases as we approach the extremal limit. At $B=0.5$, the maximum allowable spin is 0.990197 (very close to 0.99) with high power output. For $B = 0.4$, there is a significant difference between $a = 0.996273$ and $a = 0.99$, indicating that it is not very close to the extreme black hole, so the power is relatively small.

Finally, we compare the energy extraction power among Kerr-Bertotti-Robinson black holes, Kerr-Melvin black holes, and Kerr black holes in Table \ref{tab:full_power_ratio}. Here $P$, $P_{\mathrm{Kerr}}$, and $P_{\mathrm{Kerr-Melvin}}$ denote the power for Kerr-Bertotti-Robinson, Kerr, and Kerr-Melvin black holes respectively. Parameters are set to $\xi=\pi/12$ and $w=0.001$. The metric for Kerr-Melvin black holes can be found in references \cite{7,8,9,11,6,10,12}.

\begin{figure}[!h]
  \centering
  \begin{subfigure}{0.49\textwidth}
    \centering
    \includegraphics[width=\linewidth]{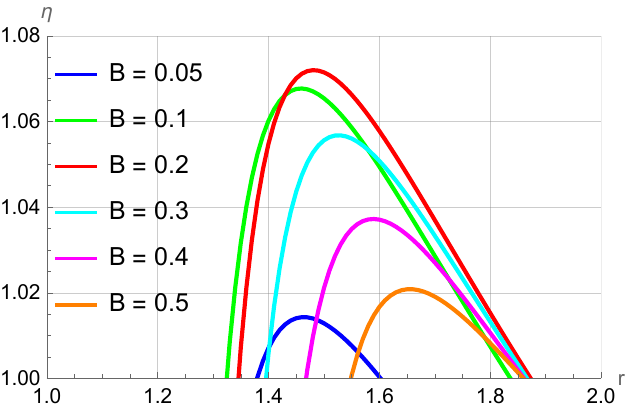}
    \caption{a=0.97} 
    \label{fig:6a}
  \end{subfigure}
  \begin{subfigure}{0.49\textwidth}
    \centering
    \includegraphics[width=\linewidth]{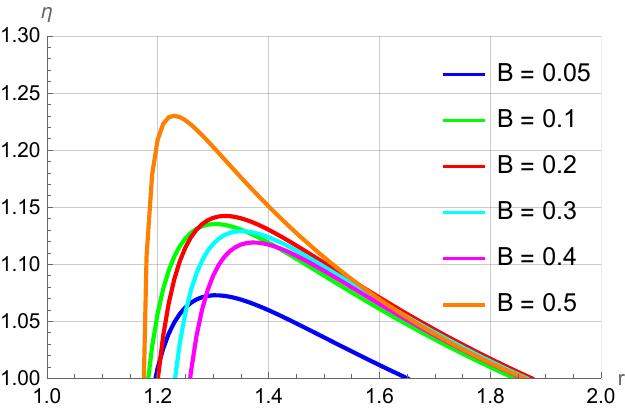} 
    \caption{a=0.99} 
    \label{fig:6b}
  \end{subfigure}
  \caption{Variation of energy extraction efficiency versus  $r$ for different $B$ values.}
  \label{fig:6}
\end{figure}

From Table \ref{tab:full_power_ratio}, it can be observed that the power of Kerr-Bertotti-Robinson black holes consistently falls below that of Kerr black holes. This discrepancy widens with increasing magnetic field strength. The difference from the Kerr black hole is somewhat smaller when $a = 0.99$. An anomalous behavior emerges at $a=0.99$ for $B = 0.45$ and $0.5$, where the power suddenly increases. This reversal stems from proximity to extremal black holes, which facilitates energy extraction, however  still fails to surpass Kerr black hole performance. This indicates that magnetic fields hinder energy extraction, with stronger fields imposing greater suppression, consistent with findings in reference \cite{6}. Notably, while magnetic fields still impede power output, the novel Kerr-Bertotti-Robinson solution enables more efficient energy extraction than Kerr-Melvin black holes. As $B$ increases, Kerr-Bertotti-Robinson's power advantage over Kerr-Melvin grows, demonstrating its superior energy-extraction capability in strong magnetic fields. Compared to the $a=0.99$ case, Kerr-Bertotti-Robinson black holes at $a=0.97$ exhibit even greater power superiority over Kerr-Melvin solutions. This suggests that away from extremal limits, Kerr-Bertotti-Robinson black holes offer enhanced advantages for energy extraction.
\begin{table}[htbp]
\centering
\small
\setlength{\tabcolsep}{4pt}
\begin{tabular}{c|cccccccccc}
\hline
\textbf{spin} & \multicolumn{10}{c}{ B} \\
\cline{2-11}
$a$ & 0.05 & 0.1 & 0.15 & 0.2 & 0.25 & 0.3 & 0.35 & 0.4 & 0.45 & 0.5 \\
\hline
\multicolumn{11}{c}{$P_{\mathrm{}} / P_{\mathrm{Kerr}}$} \\
\hline
0.97 & 0.93231 & 0.92220 & 0.84883 & 0.75473 & 0.64827 & 0.53629 & 0.42742 & 0.32510 & 0.23759 & 0.17185 \\
0.99 & 0.98029 & 0.94940 & 0.89698 & 0.83138 & 0.76111 & 0.69066 & 0.63089 & 0.59247 & 0.60993 & 0.98452 \\
\hline
\multicolumn{11}{c}{$P_{\mathrm{}} / P_{\mathrm{Kerr-Melvin}}$} \\
\hline
0.97 & 1.07047 & 1.05207 & 1.10893 & 1.22099 & 1.45030 & 2.05957 & 6.65930 & \text{none} & \text{none} & \text{none} \\
0.99 & 1.01673 & 1.03586 & 1.07957 & 1.15635 & 1.29245 & 1.55089 & 2.14975 & 4.48187 & \text{none} & \text{none} \\
\hline
\end{tabular}
\caption{Comparison of energy extraction power ratios among the three black holes in circular orbits. }
\label{tab:full_power_ratio}
\end{table}

Next, we plot the energy extraction efficiency. The efficiency of energy extraction is given by \cite{3}
\begin{align}
\eta=\frac{e_{+}^{\infty}}{e_{+}^{\infty}+e_{-}^{\infty}}.\label{38}
\end{align} 
Note that since energy extraction requires $e_{-}^{\infty}<0$ and $e_{+}^{\infty}>0$, the efficiency defined by the above expression is meaningful only when $\eta > 1$. In Fig. \ref{fig:6}, we show the variation of energy extraction efficiency with $r$ for different $B$ values, with fixed parameters $w=0.001$ and $\xi=\pi/12$. Blue, green, red, cyan, magenta, and orange curves correspond to $B$ = 0.05, 0.1, 0.2, 0.3, 0.4, 0.5, respectively. The left panel displays results for $a=0.97$, while the right panel shows $a=0.99$.

From Fig. \ref{fig:6}, we observe that efficiency first increases then decreases with $r$. As $B$ increases, efficiency initially rises then falls, again indicating that a moderate $B$ value is more favorable for energy extraction. An anomalous case occurs at $a = 0.99$ and $B = 0.5$, further demonstrating that proximity to extremal black holes facilitates energy extraction. The higher spin parameter in the right panel ($a=0.99$) results in greater efficiency compared to the left panel ($a=0.97$).

Within the circular orbital region, we conclude that either a moderate $B$ value or proximity to an extremal black hole enhances energy extraction. While magnetic fields generally impede energy extraction, the Kerr-Bertotti-Robinson black hole nevertheless exhibits stronger energy extraction capability compared to Kerr-Melvin black holes.

\section{Energy Extraction from Kerr-Bertotti-Robinson Black Holes in the Plunging Region}

\subsection{Magnetic Reconnection Process in the Plunging Region}

Next, we consider the scenario in the plunging region. Plasma initially follows a circular trajectory just outside ISCO. However, circular orbits become unstable within the ISCO radius. Consequently, any disturbance triggers radial motion, causing the plasma to plunge inwards starting precisely at the ISCO. This inner domain is termed the plunging region. Here, the presence of significant radial velocity invalidates the Keplerian velocity formula given in \eqref{27}, as that expression accounts solely for the azimuthal component.  For comprehensive computational details regarding the plunging region, refer to reference \cite{13}. Here, we briefly review the energy extraction process in the plunging region.
We continue to utilize the convenient ZAMO frame. The transformation of four-velocity components between the Boyer-Lindquist (BL) coordinates and the ZAMO frame is expressed as
\begin{align}
\hat{U}^{\mu} &= \hat{\gamma}_s \begin{pmatrix} 
1 \\ 
\hat{v}_s^{(r)} \\ 
0 \\ 
\hat{v}_s^{(\varphi)} 
\end{pmatrix}
= \begin{pmatrix} 
\dfrac{E - \omega^{\varphi} L}{\alpha} \\ 
\sqrt{g_{rr}} \, U^{r} \\ 
0 \\ 
\dfrac{L}{\sqrt{g_{\varphi\varphi}}} 
\end{pmatrix}, \label{39}
\end{align}
where 
\begin{align}
\left(U^{r}\right)^{2}=\left(\frac{d r}{d \tau}\right)^{2},   
\end{align}
where $\left(\frac{d r}{d \tau}\right)^{2}$ satisfies \eqref{15}.
Within the plunging region, the energy $E$ and angular momentum $L$ remain conserved quantities. We denote their values at ISCO as
\begin{align}
 L_{I}=L\left(r_{I}\right),E_{I}=E\left(r_{I}\right),\label{41}
\end{align}
where $ r_{I}$ is the radius of ISCO, and $E$ and $L$ satisfy \eqref{21}.
Substituting \eqref{41} into \eqref{15} , and using  $\varepsilon=1$, we get
\begin{equation}
U^{r}=-\frac{ \Omega^2 }{ r^2 } \sqrt{ \left[E_I (r^2 + a^2) - a L_I\right]^2 - \Omega^2 (a E_I - L_I)^2 \Delta - \Delta r^{2}}.\label{42}
\end{equation}
The negative sign indicates inward motion. Substituting \eqref{42} into \eqref{39} gives $\hat{v}_{s}^{(r)}, \hat{v}_{s}^{(\varphi)}$, and\\ $\hat{v}_{s}=\sqrt{\left(\hat{v}_{s}^{(r)}\right)^{2}+\left(\hat{v}_{s}^{(\varphi)}\right)^{2}}$, 
 $\hat{\gamma}_{s}$ is the Lorentz factor of $\hat{v}_{s}$ , $r_{I}$ satisfies \eqref{19}. In this case, $e_{ \pm}^{\infty}$  becomes as \cite{3,13}
\begin{align}
\begin{aligned}
e_{ \pm}^{\infty}&=\alpha \hat{\gamma}_{s} \gamma_{{out }}\left[\left(1+\beta^{\varphi} \hat{v}_{s}^{(\varphi)}\right) \pm v_{ {out }}\left(\hat{v}_{s}+\beta^{\varphi} \frac{\hat{v}_{s}^{(\varphi)}}{\hat{v}_{s}}\right) \cos \xi \mp v_{ {out }} \beta^{\varphi} \frac{\hat{v}_{s}^{(r)}}{\hat{\gamma}_{s} \hat{v}_{s}} \sin \xi\right]\\&-\alpha \left( 4 \hat{\gamma}_{s} \gamma_{{out}} \left(1 \pm \hat{v}_{s} v_{{out}} \cos \xi\right) \right)^{-1},
\end{aligned}
\end{align}
where $v_{ {out }}$ represents the outflow velocity and $\gamma_{ {out }}$ denotes its corresponding Lorentz factor. These quantities are defined by the magnetization parameter $\sigma$ as
\begin{equation}
v_{{out}}=\sqrt{\frac{\sigma}{\sigma+1}}, \gamma_{out }=\sqrt{1+\sigma}.
\end{equation}
\begin{figure}[!h]
  \centering
  \begin{subfigure}{0.32\textwidth}
    \centering
    \includegraphics[width=\linewidth]{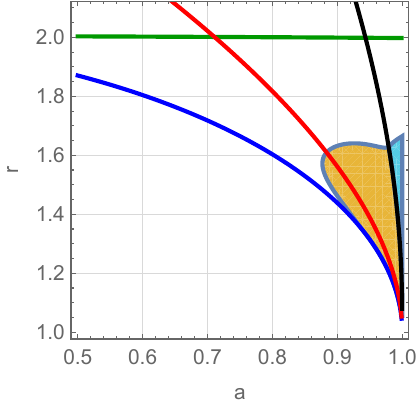}
    \caption{$B$=0.05} 
    \label{fig:7a}
  \end{subfigure}
  \begin{subfigure}{0.32\textwidth}
    \centering
    \includegraphics[width=\linewidth]{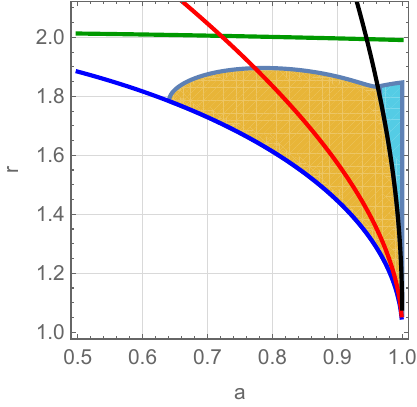} 
    \caption{$B$=0.1} 
    \label{fig:7b}
  \end{subfigure}
 \begin{subfigure}{0.33\textwidth}
    \centering
    \includegraphics[width=\linewidth]{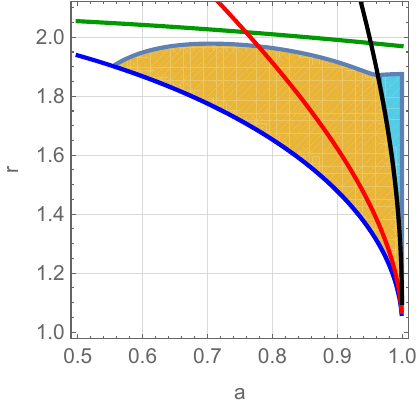} 
    \caption{$B$=0.2} 
    \label{fig:7c}
  \end{subfigure}
 \begin{subfigure}{0.32\textwidth}
    \centering
    \includegraphics[width=\linewidth]{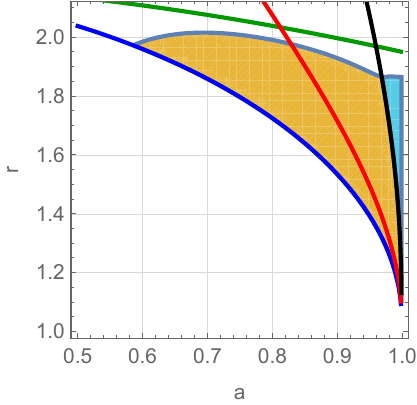} 
    \caption{$B$=0.3} 
    \label{fig:7d}
  \end{subfigure}
 \begin{subfigure}{0.32\textwidth}
    \centering
    \includegraphics[width=\linewidth]{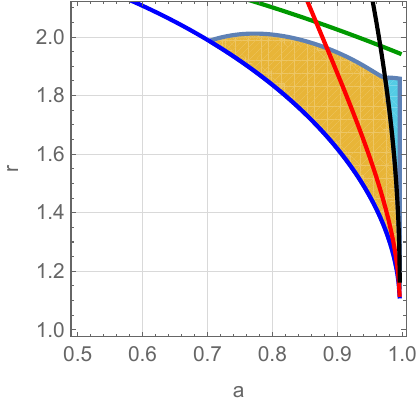} 
    \caption{$B$=0.4} 
    \label{fig:7e}
  \end{subfigure}
 \begin{subfigure}{0.32\textwidth}
    \centering
    \includegraphics[width=\linewidth]{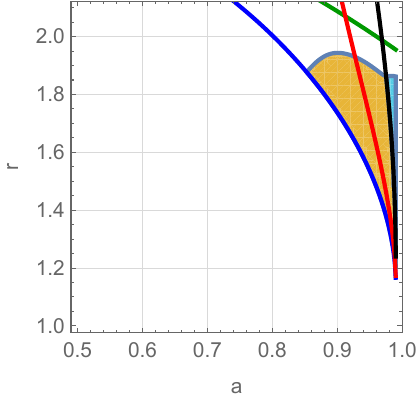} 
    \caption{$B$=0.5} 
    \label{fig:7f}
  \end{subfigure} 
  \begin{subfigure}{0.32\textwidth}
    \centering
    \includegraphics[width=\linewidth]{4_g_.pdf} 
    \caption{Legend} 
    \label{fig:4f}
  \end{subfigure} 
  \caption{Energy extraction allowable regions in the plunging region under different values of $B$. }
  \label{fig:7}
\end{figure}
Next, we plot the parameter space for energy extraction in the plunging region on the $r$-$a$ plane, shown in Fig. \ref{fig:7}. Here, blue, red, black, and green curves represent the event horizon, photon sphere radius, ISCO, and boundary of ergosphere respectively, with parameters set to $w=0.001$ and $\xi=\pi/12$. The yellow region lies within the plunging region, while the blue area still corresponds to circular orbits.

From Fig. \ref{fig:7}, we observe that as $B$ increases, the area of the allowed energy extraction  region in the plunging region first expands and then contracts. This again reinforces the conclusion that a moderate $B$ value is optimal for energy extraction.

\subsection{Power and Efficiency of Energy Extraction in the Plunging Region}
We now plot the energy extraction power in the plunging region for $a=0.97$ and $a=0.99$, employing the same formulas \eqref{31} and \eqref{32} with fixed parameters $\xi=\pi/12$ and $w=0.001$. The left panel shows $a=0.97$, while the right panel displays $a=0.99$. Blue, green, red, cyan, magenta, and orange curves correspond to $B=0.05, 0.1, 0.2, 0.3, 0.4, 0.5$ respectively.
\begin{figure}[!h]
  \centering
  \begin{subfigure}{0.49\textwidth}
    \centering
    \includegraphics[width=\linewidth]{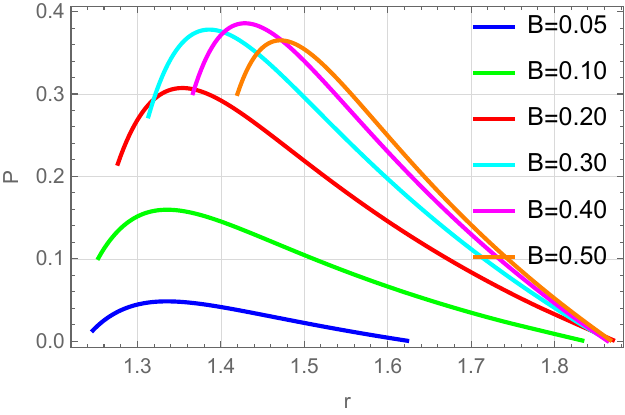}
    \caption{a=0.97} 
    \label{fig:8a}
  \end{subfigure}
  \begin{subfigure}{0.49\textwidth}
    \centering
    \includegraphics[width=\linewidth]{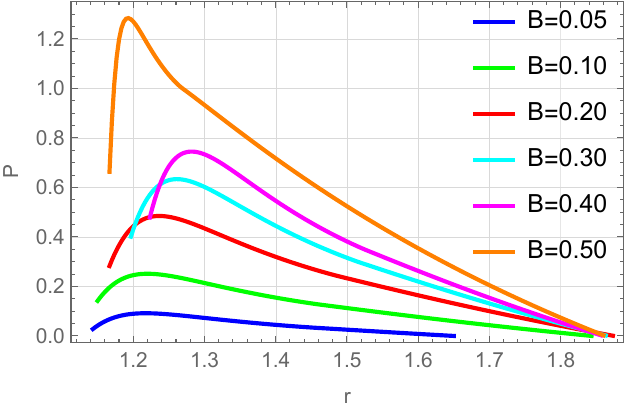} 
    \caption{a=0.99} 
    \label{fig:8b}
  \end{subfigure}
  \caption{Variation of energy extraction per unit enthalpy power versus  $r$ in the plunging region for different $B$ values. }
\label{fig:8}
\end{figure}

From Fig. \ref{fig:8}, we observe that in the plunging region, the trend of power variation with increasing $B$ closely mirrors that in the circular orbit region. This further confirms that either a moderate $B$ value or proximity to an extremal black hole enhances energy extraction.

Within the plunging region, we similarly compare the power outputs among Kerr-Bertotti-Robinson, Kerr, and Kerr-Melvin black holes in Table \ref{tab:plunge_power_ratio_full}. First, comparing Kerr and Kerr-Bertotti-Robinson black holes. Table \ref{tab:plunge_power_ratio_full} reveals that magnetic fields still impede energy extraction, with stronger fields causing greater suppression. However, the power ratios here are more favorable than in circular orbits, except for $a=0.99$ at $B=0.5$. The anomalous rise at $a=0.99$ for $B=0.45$ and $0.5$ stems from extremal black hole effects, with both plunging and circular orbit regions exhibiting increased power near $B=0.45$. This reaffirms that proximity to extremal black holes facilitates energy extraction. Next, comparing Kerr-Bertotti-Robinson and Kerr-Melvin black holes. The variation pattern in the plunging region resembles that in circular orbits. The novel Kerr-Bertotti-Robinson solution consistently outperforms Kerr-Melvin black holes, demonstrating greater advantages in strong magnetic fields and particularly when not near extremal limits. 
\begin{table}[htbp]
\centering
\small
\setlength{\tabcolsep}{4pt}
\begin{tabular}{c|cccccccccc}
\hline
\textbf{spin} & \multicolumn{10}{c}{ B } \\
\cline{2-11}
$a$ & 0.05 & 0.1 & 0.15 & 0.2 & 0.25 & 0.3 & 0.35 & 0.4 & 0.45 & 0.5 \\
\hline
\multicolumn{11}{c}{$P_{\mathrm{}} / P_{\mathrm{Kerr}}$} \\
\hline
0.97 & 0.97967 & 0.94823 & 0.89354 & 0.82306 & 0.74735 & 0.66432 & 0.58258 & 0.50590 & 0.43605 & 0.38140 \\
0.99 & 0.98710 & 0.95881 & 0.91346 & 0.85841 & 0.79775 & 0.73893 & 0.68526 & 0.64839 & 0.65019 & 0.89167 \\
\hline
\multicolumn{11}{c}{$P_{\mathrm{}} / P_{\mathrm{Kerr-Melvin}}$} \\
\hline
0.97 & 1.01261 & 1.02572 & 1.05277 & 1.10345 & 1.19444 & 1.35971 & 1.72982 & \text{none} & \text{none} & \text{none} \\
0.99 & 1.00990 & 1.02403 & 1.05263 & 1.10478 & 1.18580 & 1.32914 & 1.59259 & 2.19764 & \text{none} & \text{none} \\
\hline
\end{tabular}
\caption{Comparison of energy extraction power ratios among three black hole types in the plunging region. }
\label{tab:plunge_power_ratio_full}
\end{table}

Finally, we examine energy extraction efficiency in scenarios including plunge trajectories for $a=0.97$ and $a=0.99$. Employing the same efficiency formula \eqref{38} with fixed parameters $w=0.001$ and $\xi=\pi/12$, we plot the results where blue, green, orange, magenta, cyan, and red curves correspond to $B=0.05, 0.1, 0.2, 0.3, 0.4, 0.5$ respectively.
\begin{figure}[!h]
  \centering
  \begin{subfigure}{0.49\textwidth}
    \centering
    \includegraphics[width=\linewidth]{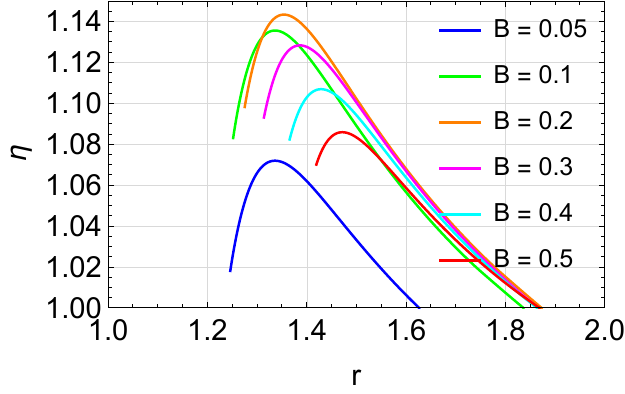}
    \caption{a=0.97} 
    \label{fig:9a}
  \end{subfigure}
  \begin{subfigure}{0.49\textwidth}
    \centering
    \includegraphics[width=\linewidth]{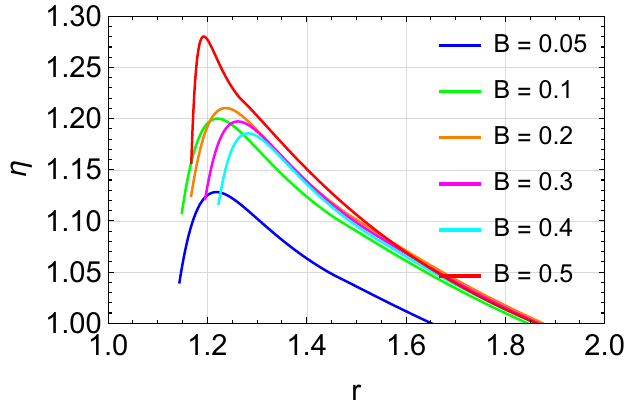} 
    \caption{a=0.99} 
    \label{fig:9b}
  \end{subfigure}
  \caption{Efficiency of energy extraction in the plunging region versus  $r$ for different $B$ values. }
\label{fig:9}
\end{figure}
From Fig. \ref{fig:9}, it can be seen that as $B$ increases, the trend of efficiency variation in the plunging region remains similar to that in circular orbits. This further indicates that either a moderate $B$ value or proximity to an extremal black hole is more favorable for energy extraction.

In summary, within the plunging region, energy extraction is also optimized by either a moderate $B$ value or closeness to an extremal black hole. While the magnetic field still impedes energy extraction, the energy extraction capability here surpasses that in circular orbits. The Kerr-Bertotti-Robinson black hole consistently demonstrates stronger energy extraction capacity than the Kerr-Melvin black hole.
\section{Conclusion}

Recent years, more and more astronomical observations indicates the presence of magnetic fields around black holes. Black hole solutions incorporating magnetic fields are therefore crucial for understanding significant astrophysical phenomena. Existing descriptions of magnetized black holes, such as Melvin-like black holes, has many drawbacks.  For example,  magnetic fields diminish at large distances,  geodesics in these spacetimes cannot escape to infinity, and their algebraic type is $I$, rendering them unrealistic models for astrophysical magnetized black holes. The exact solution recently proposed in Ref. \cite{1} circumvents these issues. Many properties of this black hole remain unexplored. So we investigated energy extraction from the Kerr-Bertotti-Robinson black hole via the Comisso-Asenjo process and compared the results with those of the Kerr-Melvin   black hole. Specifically, we conducted studies in both the circular orbits region and the plunging region and compared the corresponding results.

First, we provided a brief introduction to the spacetime of the Kerr-Bertotti-Robinson black hole. We plotted the size of the ergoregion versus $B$ and found that near extremality, the ergosphere widens, while increasing $B$ narrows the size of the ergoregion, hindering energy extraction. This indicates that the magnetic field impedes energy extraction. We then plotted the event horizon and boundary of ergosphere for different $B$ values versus $a$, aiding the understanding of ergosphere variations. Finally, we analyzed energy extraction from the Kerr-Bertotti-Robinson  black hole in circular orbits. Besides explaining the energy conversion mechanism via magnetic reconnection in circular orbits and plotting energy extraction parameter diagrams, we studied the power and efficiency of energy extraction in this region. We discovered that either a moderate $B$ value or proximity to an extremal black hole facilitates more effective energy extraction. Comparing the energy extraction power with that of a Kerr black hole revealed that the magnetic field hinders extraction, with stronger fields causing greater hindrance, a finding similar to that observed in the Melvin-like black hole. Crucially, direct comparison with the  Kerr-Melvin  black hole showed that the energy extraction power is stronger in the Kerr-Bertotti-Robinson  spacetime. We also investigated the magnetic reconnection mechanism and energy extraction in the plunging region. There too, a moderate $B$ value or closeness to extremality was found to be more favorable for energy extraction, and the magnetic field still impeded the process,  which are identical to those for the circular orbits region. Similarly, in the plunging region, we compared the energy extraction power between the Melvin-like and Kerr-Bertotti-Robinson  black holes, and the results demonstrate that the Kerr-Bertotti-Robinson black hole retains its advantage.

Comparing the results from the circular orbits and plunging regions, we found that for all the black hole, the power and efficiency of energy extraction are higher in the plunging region than in the circular orbits region. This finding is entirely consistent with the conclusion in Ref. \cite{38}.

\noindent {\bf Acknowledgments}

\noindent
This work is supported by the National Natural Science Foundation of China (Grants No. 12375043).

\end{document}